\documentclass[twocolumn]{revtex4}
\usepackage{aas_macros}
\usepackage{color}
\usepackage{epsfig,graphicx}
\usepackage{graphics}
\usepackage{latexsym,amssymb}
\usepackage{amssymb}
\usepackage{amsmath}
\begin{document}

\title{The CMB in a Causal Set Universe}
\author{Joe Zuntz}
\affiliation{Astrophysics, Denys Wilkinson Building, Keble Road, Oxford OX1 3RH, UK}
\date{November 19, 2007}

\begin{abstract}
We discuss Cosmic Microwave Background constraints on the causal set theory of quantum gravity, which has made testable predictions about the nature of dark energy.  We flesh out previously discussed heuristic constraints by showing how the power spectrum of causal set dark energy fluctuations can be found from the overlap volumes of past light cones of points in the universe.  Using a modified Boltzmann code we put constraints on the single parameter of the theory that are somewhat stronger than previous ones.  We conclude that causal set theory cannot explain late-time acceleration without radical alterations to General Relativity.
\end{abstract}
\maketitle

\newcommand{\evl}{Everpresent $\lambda$}
\newcommand{\cl}{\ensuremath{C_\ell}}

\section{Introduction}
Causal sets or causets are an approach to quantum gravity in which spacetime is discretized with causality as the fundamental relation between the discrete elements.  The requirement that the as-yet unknown theory of causet dynamics should give rise to a smooth classical spacetime imposes strong restrictions on the character of the theory.  In particular, the requirement that physics should appear to be Lorentz-covariant implies a Poisson relation between the classical volume of a region of spacetime and the number of discrete elements in it.  The mean number of elements per Planck volume, $\alpha^2$, in the Poisson process is the only free parameter in the theory on macroscopic scales.

It was noted in \citet{forksintheroad} that the fluctuations in the Poisson process naturally give an appropriately sized dark energy if the volume of the universe is taken as conjugate to $\Lambda$.  In \citet{evl} it was shown that the same principle predicts a randomly fluctuating $\Omega_\Lambda$ of roughly constant amplitude $\alpha$ throughout the history of the universe, with the name everpresent $\Lambda$ given to the phenomenon.

A fundamental objection to this explanation of dark energy was raised in \citet{barrow}, which noted that since everpresent $\Lambda$ is sourced in past light cones (PLCs) of points, regions of the last-scattering surface separated by more than a causal horizon should have uncorrelated $\Lambda$, and should produce $o(\alpha)$ CMB anisotropies.  A limit of $\sim 10^{-5}$ was therefore placed on $\alpha$, too small for it to explain late-time acceleration.  In this paper we flesh out the objection in \citet{barrow} with a numerical analysis based on the volumes of overlapping PLCs and place more constraints on $\alpha$.

To emphasize that we are not considering a cosmological constant we will break from the notation used in \citet{evl} and use $\lambda$ to represent the causet dark energy, rather than $\Lambda$.

\section{Theory}

We make use of a formulation for Everpresent $\lambda$ introduced in \citet{evl}:  we consider the $\lambda$ to be the total Hamiltonian action $S$ per unit spacetime volume $V$, with each causet element having randomly signed action $\pm \alpha^2$:
\begin{equation}
\lambda = S  / V
\label{eq-lam-s-v}
\end{equation}

The volume $V$ is of the past light cone (PLC) of any point, and in a discrete spacetime is proportional to the number of elements $N$ in it.  In conformal time $\tau$, $V$ is given by
\begin{eqnarray}
V(\tau) &\equiv& \alpha^2 N   \\ 
&=& \frac{4\pi}{3} \int_0^\tau (\tau-\tau')^3 a^4(\tau') \mathrm{d}\tau'\, .
\label{eq-volume}
\end{eqnarray}

Since the action for each element is randomly chosen as $\pm \alpha^2$, the total comes from binomial statistics:
\begin{eqnarray}
\langle S \rangle &=& 0 \, , \nonumber  \\
\langle S^2 \rangle &=& \alpha^4 N \, .
\end{eqnarray}
The variance of $\lambda$ is therefore given by
\begin{equation}
\langle \lambda^2 \rangle = \frac{1}{N} = \frac{\alpha^2}{V}\, .
\end{equation}
Since $V^{-\frac{1}{2}} \propto H^2 \propto \rho_c$ this formulation provides a density which oscillates randomly with amplitude $\alpha \rho_{\textrm{crit}}$  at every epoch --- an everpresent $\lambda$.

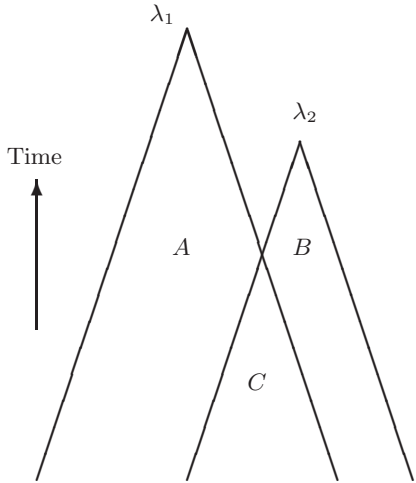
\begin{figure}[h]
\setlength{\unitlength}{2cm}
\centering
\thicklines
\begin{picture}(3,3)
\put(0,0){\line(1,3){1}}
\put(2,0){\line(-1,3){1}}

\put(1,0){\line(1,3){0.75}}
\put(2.5,0){\line(-1,3){0.75}}
\put(0,1){\vector(0,1){1}}
\put(0.75,3.05){$\lambda_1$}
\put(1.7,2.4){$\lambda_2$}
\put(0.9,1.5){$A$}
\put(1.7,1.5){$B$}
\put(1.4,0.6){$C$}
\put(-0.2,2.1){Time}
\end{picture}

\caption[Past light cone overlaps schematic]{\label{cau-fig-inhom-schem} A schematic of the PLCs of two points.  The Big Bang is at the bottom of the diagram; the dimensions are heavily distorted.  The volumes of the light cones  are $V_1=V_A+V_C$ and $V_2=V_B+V_C$.} \label{fig-schematic}
\end{figure}

We can also use this formulation to determine the two-point correlations of $\lambda$.  Since $\lambda$ is sourced by the action in a PLC, we split the PLCs of two points into shared and disjoint regions.  We use the notation $\lambda_A\equiv \lambda(x_A)$ and consider the volume splitting that is illustrated in figure \ref{fig-schematic}.
\begin{eqnarray}
\textrm{corr}(\lambda_1,\lambda_2)  & \equiv & \frac{\langle \lambda_1 \lambda_2 \rangle}{\sqrt{\langle\lambda_1^2\rangle\langle\lambda_2^2\rangle}} \nonumber \\
&=& \frac{\langle(S_A+S_C)(S_B+S_C)\rangle}{\sqrt{\langle S_1^2 \rangle \langle S_2^2 \rangle }  } \nonumber \\
&=& \frac{\langle S_C^2 \rangle}{\sqrt{\langle S_1^2 \rangle \langle S_2^2 \rangle}} \nonumber \\
&=& \frac{V_C}{\sqrt{V_1 V_2}} \label{eq-corr-overlap}
\end{eqnarray}
The second line follows because the volumes cancel, the third because disjoint volumes have zero correlation, and the fourth by construction of $S$.

We specialize to the case where the points' PLC volumes $V_1$ and $V_2$  are equal --- the points are in a statistically homogenous universe and have the same time co-ordinate.  At any point in  the history of the universe the instantaneous three-volume of the overlapping region is the overlap between the two points' (spherical) horizon volumes.  The 4-volume can then be found by integrating this quantity.  The overlap volume of two spheres of equal radii $r$ and separation $d$ is \citep{mathworld-spheresphere}:
\begin{equation}
V^{(3)}_{o} = 
\frac{\pi}{12}(4r+d)(2r-d)^2
\end{equation}
if $d<2r$, and zero otherwise.

In our case for a point at time $\tau$, the horizon of a point a time $\tau'$ earlier grown forward to $\tau$ is $(\tau-\tau')$.  For a comoving separation $\chi$ the proper separation is $a(\tau)\chi$ and the comoving horizon is $\tau$, so converting to a proper 3-volume:
\begin{equation}
V^{(3)}_o(\tau,\tau',\chi) =
a^3(\tau)\frac{\pi}{12}\left( 4(\tau-\tau')+\chi\right)\left(2(\tau-\tau')-\chi\right)^2
\end{equation}
if $\chi<2(\tau-\tau')$, and zero otherwise.

To compute the 4-volume of the PLC at $\tau$ we integrate this quantity with the time element $a\mathrm{d}\tau$:
\begin{eqnarray}
V_{o}(\tau,\chi) = \frac{\pi}{12}\int_0^{\tau-\chi/2}  \!\!\!\!\! &&a^4(\tau)\left( 4(\tau-\tau')+\chi\right) \nonumber \\
&&\cdot \left(2(\tau-\tau')-\chi\right) \mathrm{d}\tau' \, .
\label{cau-eq-overlap}
\end{eqnarray}

We use the $V_1$ and $V_2$ from equation \ref{eq-volume}, and get the covariance by multiplying by $\sqrt{\langle \lambda^2 \rangle } = 8\pi G \alpha \bar{\rho}$. 

To obtain CMB spectra from the $\lambda$ overlap we require a source term that converts from metric fluctuation to radiation anisotropy (see \citet{SELZAL}).  We use:
\begin{equation}
S_T(k,\tau) = g\psi\, , 
\label{cau-eq-sources}
\end{equation}
where $g=\dot{\kappa}e^{-\kappa}$ is the visibility function, $\kappa$ is the optical depth to time $\tau$, and $\psi$ is the conformal gauge perturbation variable.  $g$ is computed in a standard way from the cosmic reionization history.  We have neglected the effect of the metric fluctuation on the plasma, since the latter does not have time to respond to the former, which changes over a Hubble time.  We have also neglected the Integrated Sachs-Wolfe effect.  We can find an expression for $\psi$ from the Einstein equations, but it will always be in terms of unknown stress-energy components.  

We have no stress-energy tensor for $\lambda$ that we could use to fully compute $\psi$ --- we only have one component.  We will discuss two assumptions that can be made about $\lambda$ dynamics to make progress.  Implementations of each are discussed in section \ref{sec-imp}.

\subsection{Zero Velocity Assumption}
If we substitute for the spatial conformal perturbation $\phi$ in the perturbed Einstein equations \citep{MaB} we can obtain $\psi$.  Letting:
\begin{eqnarray}
D \equiv 4 \pi G a^2 \delta T^0_0 = a^2 \lambda/2 \nonumber \, ,  \\
\Upsilon \equiv i 4 \pi G  a^2  \delta k^jT^0_j = \mbox{unkown} \, , 
\end{eqnarray}
then:
\begin{eqnarray}
k^2(\dot{\phi} + H\psi) = \Upsilon \nonumber \, ,  \\ 
k^2 \phi + 3H(\dot{\phi}+H\psi) = D \, . 
\end{eqnarray}

One approach is to neglect $\Upsilon$ altogether, and have the CMB sourced by $\lambda$ alone.  This might be expected to give an answer of the correct order, since velocity and density perturbations are typically the same magnitude, and unlikely to cancel exactly.  In that case:
\begin{eqnarray}
\psi &=& - H^{-1} \dot{\phi} \nonumber \\
&=& - k^{-2} H^{-1} \dot{D} \nonumber \\
&=& -k^{-2} H^{-1} \left(\dot{a} a \lambda + a^2 \dot{\lambda}\right) \nonumber \\
&=& -k^{-2} a^2 \lambda \, .
\end{eqnarray}
where in the last line we have set $\dot{\lambda}=0$ because it is formally undefined.

In this approximation we therefore take $\psi_k = -a^2 \lambda_k / k^2$.  

\subsection{Zero Shear Assumption}

Another reasonable assumption that we could make would be that like many other forms of dark energy, the \evl\ has zero anisotropic metric shear.  In that case $\psi=\phi$ and we can find $\psi$ in terms of $\delta T^0_0 = \lambda/8 \pi G$:

\begin{equation}
k^2 \psi + 3 H \left(\dot{\psi}+H\psi\right) =  \frac{1}{2}a^2\lambda \label{cau-eq-zeroshear1}\, . \\
\end{equation}

\begin{equation}
\Rightarrow \dot{\psi} = -\frac{k^2+3H^2}{3H}\psi+\frac{1}{6H}a^2\lambda
\end{equation}
which has the solution:
\begin{equation}
\begin{split}
\psi = \frac{1}{2}\int_0^\tau \exp\left(-\int_{\tau'}^{\tau}\frac{k^2+3H^2(\tau'')}{3H(\tau'')}\mathrm{d}\tau''\right) \\
\frac{a^2(\tau')}{H(\tau')} \lambda(\tau) \mathrm{d}\tau' \, .
\end{split}
 \end{equation}
We have set the constant of integration to zero so that there is no perturbation at zero time.  The $H$ part of the integrating factor can be split off and integrated to $\log{a}$ so that:
\begin{equation}
\psi_k = \frac{1}{2a(\tau)}\int_0^\tau Q_k(\tau,\tau') \frac{a^3(\tau')}{H(\tau')} \lambda_k(\tau') \mathrm{d}\tau'  \, , 
\label{eq-shear-psi}
\end{equation}
where:
\begin{equation}
Q_k(\tau,\tau')=\exp\left(-\frac{1}{3}k^2 \int_{\tau'}^{\tau} H^{-1}(\tau'') \mathrm{d}\tau''\right) \, .
\end{equation}
Here $Q$ is playing the role of a kernel, defining the length of time over which $\lambda$ contributes to $\psi$.  It peaks to unity at $\tau'=\tau$ and falls off over about a Hubble time, or longer for longer wavelengths.  We can compute $Q$ simply from the background dynamics of the system.

Equation \ref{eq-shear-psi} relates the dynamic $\lambda$ and $\psi$.  For our purposes we need the relation between the root-mean-square quantities.  This becomes somewhat more computationally involved:
\begin{eqnarray}
\begin{split}
\langle \psi_k^* \psi_k \rangle = \frac{1}{4a^2(\tau)}\int_0^\tau \!\! \int_0^\tau 
Q_k(\tau,\tau_1) Q_k(\tau,\tau_2)  \\
 \frac{a^3(\tau_1)}{H(\tau_1)}\frac{a^3(\tau_2)}{H(\tau_2)}  \\
 \langle \lambda_k^*(\tau_1) \lambda_k(\tau_2) \rangle  \\
 \mathrm{d}\tau_1 \mathrm{d}\tau_2 \, .
\end{split}
\label{eq-cau-zshear-psi}
\end{eqnarray}
We use the square-root of this quantity in the CMB sources.  The quantity $\langle  \lambda_k^*(\tau_1) \lambda_k(\tau_2)\rangle$ is an \emph{unequal time correlator} similar to ones used in studies of textures and strings \citep{UTC}.  We need to calculate it for each pair of times $\tau_1,\tau_2$ where $Q$ is not negligible, for each wavenumber $k$.

As with the equal time version (to which it should reduce when $\tau_1=\tau_2$), we use equation \ref{eq-corr-overlap} to compute this correlator.  This time, though, $V_1\neq V_2$.  To calculate $V_{12}$ we once again consider overlapping spheres, this time of unequal radii.  The overlap volume of two spheres of radii $r_1$ and $r_2$ with separation $d$ is \citep{mathworld-spheresphere}:
\begin{equation}
V^{(3)}_o = (r_1+r_2-d)^2(d^2+2 d (r_1+r_2) -3(r_1^2 +r_2^2) +6r_1r_2) \, .
\end{equation}

Once again we substitute $d(\tau') = a(\tau')\chi$, $r_1=a(\tau')(\tau_1-\tau')$ and $r_2=a(\tau')(\tau_2-\tau')$, and integrate:
\begin{equation}
\begin{split}
V_o(\tau_1,\tau_2) = \frac{\pi}{12}  \int_0^{(\tau_1 +\tau_2-\chi)/2}  \!\!\!\!\!\!\!\!\!\!\!\!\!\!\!\!\! & \left(   \tau _1+\tau _2-2 \tau -\chi '\right){}^2 \\
&   \left[ \chi ^2+2 (\tau _1+\tau_2 -2 \tau') \chi \right. \\
& -3\left( (\tau _1-\tau '){}^2+(\tau _2-\tau '){}^2\right)  \\ 
& \left.+6(\tau _1-\tau ')(\tau_2-\tau ') \right] \\
&  a^4(\tau') \mathrm{d}\tau' \, .
\end{split}
\label{cau-eq-unqualVolume}
\end{equation}

We get $V_1$ and $V_2$ from equation \ref{eq-volume}, and go from the correlation to the covariance by multiplying by $\sqrt{\langle\lambda^2\rangle} = 8 \pi G \alpha \bar{\rho} $.

\section{Implementation} \label{sec-imp}

We modified the CMB Boltzmann code CMBEASY \citep{CMBEASY} to include the perturbations described.  The code, which is restricted to flat universes, first computes the background evolution of the cosmos, which we leave unchanged,  and then evolves perturbations (which we skip for the causet $\lambda$), and finally computes CMB sources and calculates their \cl.  We compute the $\lambda$ contribution at the same values of $\tau$ and comoving wavelength $k$ as the code uses in its standard mode, since these are the periods when the visibility $g$ is significant enough for the perturbations to make a contribution to the CMB.

We obtain quantities in harmonic $k$-space using a Fast Fourier Transform routine \citep{fftw} from the real space quantities that we calculate using the volume overlap functions.

The expression for the PLC volume in equation \ref{eq-volume} can be split up into:
\begin{equation} 
V(\tau) = \frac{4 \pi}{3} \left[ \tau^3 I_0(\tau) - 3 \tau^2 I_1(\tau) +3\tau I_2(\tau) -I_3(\tau) \right]  \, , 
\label{cau-eq-volume2}
\end{equation}
where:
\begin{eqnarray}
I_n &=& \int_0^{\tau} \tau'^n a^4(\tau') \mathrm{d}\tau'  \, ,  \label{cau-eq_Idef} \\
\frac{\mathrm{d} I_n}{\mathrm{d} \tau} &=& \tau^n a^4(\tau) \, .  \label{cau-eq-Idot}
\end{eqnarray}

And thus:
\begin{eqnarray}
\dot{V} = & \frac{4\pi}{3} & \left(  3\tau^2 I_0 + \tau^3 \dot{I_0} - 6\tau I_1 - 3\tau^2 \dot{I_1} \right. \nonumber \\
&&\left. + 3I_2 +3\tau\dot{I_2}-\dot{I_3}\right)  \nonumber \\
&= \frac{4\pi}{3} & \left(  3\tau^2 I_0 +\tau^3 a^4 - 6\tau I_1 - 3 \tau^3 a^4 \right. \nonumber \\
&& \left. + 3I_2 + 3\tau^2 a^4 - \tau^3 a^4 \right)\nonumber \\
&= 4\pi & \left( \tau^2 I_0 -2\tau I_1 +3I_2\right)\, .
\end{eqnarray}

We can split the overlapping volumes of the PLCs in equation \ref{cau-eq-overlap} in a similar way.  The integrals range from $0$ to the time when the horizon spheres are smaller than the comoving separation of the points.  We therefore use the quantities:
\begin{eqnarray}
J_n(\tau,\chi) &=& \int_0^\tau \tau'^n a^4(\tau') \mathrm{d}\tau - \int_{\chi/2}^\tau \tau'^n a^4(\tau') \mathrm{d}\tau \nonumber \\
&=& \int_0^{\tau-\chi/2} \!\!\!\!\!\!\!\! \tau'^n a^4(\tau') \mathrm{d}\tau  \, ,
\end{eqnarray}
provided the integral range is positive.   We can use the functions $I_n$ defined in equation \ref{cau-eq-Idot} to express $J_n$:
\begin{equation}
J_n(\tau,\chi) = 
I_n ( \tau - \chi / 2)
\end{equation}
if $\tau>\chi/2$, and zero otherwise.

Using these quantities the overlap volume is:

\begin{eqnarray}
V_o=\frac{\pi}{12}\big[ &-&16 J_3 \nonumber
\\&+&12(4\tau-\chi) J_2\nonumber
\\&-&24(2\tau^2-\tau\chi) J_1 \nonumber
\\&+&(16\tau^3-12\tau^2\chi+\chi^3) J_0 \big] \label{cau-eq-overlapsplit-v} \, .
\end{eqnarray}

We use this formula, with the volume formula, to compute the $\lambda_k$ covariance for the CMB sources.

To compute the unequal-time overlaps in equation \ref{cau-eq-unqualVolume} we define yet another set of retarded volumes.  In this case the latest time $\tau$ at which the horizons overlap is a function of both $\tau_1$ and $\tau_2$:
\begin{eqnarray}
V_o=\frac{\pi}{12}\big[ &-&16 K_3 \nonumber
\\&+&12(\chi(\chi-s)+d^2)K_2 \, /\chi  \nonumber
\\&-&12((\chi-s)(s \chi-d^2)) K_1 / \, \chi \nonumber
\\&+&(\chi-s)^2 (\chi^2-3d^2+2s\chi) K_0 \, / \chi \big] \label{cau-eq-overlap2-s}  \, , 
\end{eqnarray}
where 
\begin{equation}
K_n(\chi,\tau_1,\tau_2) = 
I_n ((\tau_1+ \tau_2-\chi)/2) 
\end{equation}
if $\tau_1+\tau_2>\chi$, and zero otherwise, and 
\begin{eqnarray}
s = \tau_1+\tau_2 \nonumber  \, , \\
d = \tau_1-\tau_2 \, .
\end{eqnarray}

Since computing the overlap volumes was somewhat computationally intensive, and was repeated for each $k$, we wrote code to perform it once for all the relevant $k$ and pairs $\tau_1$ and $\tau_2$ during the peak of recombination, and save the data.  We then looked up the result in the main code where necessary. 

We use two other devices to speed the code: we compute the correlators only when the kernels $Q_k$ are not small (that is, only for $\tau_1$,$\tau_2$ in the recent past of $\tau$), and we compute the overall sources only when the visibility is greater than $1\%$ of its maximum.

\section{Results and Commentary}

Our results are in the form of standard CMB power spectra, for temperature only since we do not consider any polarization sources in the calculation.  In each case we obtain a limit on the fundamental parameter $\alpha$ of the causal set theory.

\begin{figure}[h]
\begin{center}
\includegraphics[width=8cm]{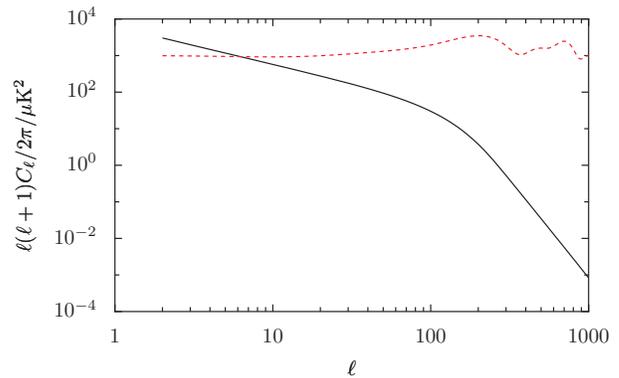}
\caption[TT spectrum for zero velocity causets]{The power spectra obtained from a standard CDM parameter combination described above (red, dashed) and from the causal set fluctuations (black), in the case of zero velocity, for $\alpha=10^{-6}$.}
\label{cau-fig-cl2}
\end{center}
\end{figure}

The result from the zero velocity assumption, for $\alpha=10^{-6}$, is shown in figure \ref{cau-fig-cl2}.  It has two expected features: power-law increase with scale, which occurs as the the separation between points on the LSS increases and their shared volume rapidly decreases; and a kink $\ell \approx 200$, which corresponds to the causal horizon on the last scattering surface, and is thus where volumes become completely disconnected.

\begin{figure}[h]
\begin{center}
\includegraphics[width=8cm]{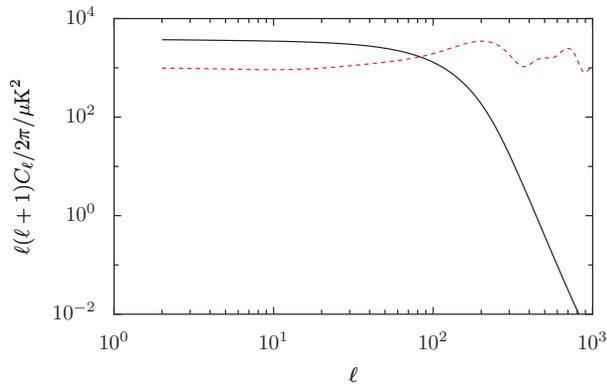}
\caption[TT spectrum for zero shear causets]{The power spectra obtained from the standard CMB parameter combination (red, dashed) and from the causal set fluctuations (black, solid), in the case of zero shear, for $\alpha=10^{-6}$.}
\label{cau-fig-cl3}
\end{center}
\end{figure}

Figure \ref{cau-fig-cl3} shows the result  of the zero shear computation, for $\alpha=10^{-7}$.  The kink at the causal horizon is retained, but the slope at large scales is reduced because at large scales the terms with the $k^2$ factor in equation \ref{cau-eq-zeroshear1} (which were implicitly all we included in the zero velocity assumption) lose dominance.

\section{Conclusions}
\label{sec-conclude}

Under both our assumptions we can place strong limits on the value of the fundamental causet parameter $\alpha$, which controls the amplitude of all the causet fluctuations and thus (quadratically) controls the power spectra scales.  In both cases, and probably in all simple models, the causet $C_\ell$ are largest at low $\ell$.  This makes comparing the predictions to data straightforward; we can simply see at what $\alpha$ the low-$\ell$ causet \cl\ is equal to the measured value.  A higher $\alpha$ would lead to higher \cl\  and is therefore ruled out.

In both cases, as predicted by Barrow, the large scale CMB values limit $\alpha$ to well below $1$.  The exact constraint depends on the detail of the conversion from the $\lambda$ perturbation to $\psi$.  In the case of the zero velocity assumption described above the limit is is approximately $\alpha < 10^{-6}$.  In the case of the zero shear assumption the limit is approximately $\alpha < 10^{-7}$.  In both these cases, again as Barrow notes, the given $\alpha$ is too small to generate late time acceleration.  We would require $\alpha$ to be at least  $o(10^{-2})$ to achieve that.  We therefore concur with Barrow and find no loophole in the mechanics of causal set structure formation that  would invalidate his conclusion that \evl\ is not dark energy.

From a positive perspective, we can think of these results as providing direction to the theory.  A successful version of it must resolve the problem of excess structure formation, but retain acceleration at late times.  This might come about by finding a very unusual stress-energy tensor, by suppressing $\lambda$ at early times, or by modifying the Einstein equations more dramatically.  The last seems the most promising possibility.

\begin{acknowledgements}
I am grateful to Rafael Sorkin and Fay Dowker for an introduction to, and useful discussions about, causets.  I am also grateful to Andrew Jaffe for much useful discussion about causal sets and many other things.  Most of this work was carried out during an STFC studentship; some was also carried out during a visit to the Perimeter Institute for Theoretical Physics. 
\end{acknowledgements}

\end{document}